\documentclass[lineno]{jfm}

\usepackage{graphicx}
\usepackage{newtxtext}
\usepackage{newtxmath}
\usepackage{natbib}
\usepackage{hyperref}
\hypersetup{
    colorlinks = true,
    urlcolor   = blue,
    citecolor  = black,
}

\newcommand{\RomanNumeralCaps}[1]
\linenumbers

\usepackage{caption}
\newcommand{\tabincell}[2]{\begin{tabular}{@{}#1@{}}#2\end{tabular}}

\shorttitle{Sea Water Freezing Modes in a Natural Convection System}
\shortauthor{Y. Du, Z. Wang, L. Jiang, E. Calzavarini and C. Sun }

\author{Yihong Du\aff{1}\thanks{Equally contributed authors},
  Ziqi Wang\aff{1}\thanks{Equally contributed authors}, Linfeng Jiang\aff{1}, Enrico Calzavarini\aff{2}\corresp{\email{enrico.calzavarini@univ-lille.fr}}
 \and Chao Sun\aff{1,3}\corresp{\email{chaosun@tsinghua.edu.cn}}}

\affiliation{\aff{1}Center for Combustion Energy, Key Laboratory for Thermal Science and Power Engineering of MoE, International Joint Laboratory on Low Carbon Clean Energy Innovation, and Department of Energy and Power Engineering, Tsinghua University, 100084 Beijing, China
\aff{2}{Univ.\ Lille, Unit\'e de M\'ecanique de Lille - J. Boussinesq (UML) ULR 7512, F-59000 Lille, France}
\aff{3}{Department of Engineering Mechanics, School of Aerospace Engineering, Tsinghua University, Beijing 100084, China}}

\title{Sea Water Freezing Modes in a Natural Convection System}

\begin{document}

\maketitle

\begin{abstract}
Sea ice is crucial in many natural processes and human activities. Understanding the dynamical couplings between the inception, growth and equilibrium of sea ice and the rich fluid mechanical processes occurring at its interface and interior is of relevance in many domains ranging from geophysics to marine engineering. Here we experimentally investigate the complete freezing process of water with dissolved salt in a standard natural convection system, i.e., the prototypical Rayleigh-B\'enard cell. Due to the presence of a mushy phase, the studied system is considerably more complex than the freezing of freshwater in the same conditions~\citep{wang2021growth}. 
\textcolor{black}{We measure the ice thickness and porosity at the dynamical equilibrium state for different initial salinities of the solution and temperature gaps across the cell. These observables are non-trivially related to the controlling parameters of the system as they depend on the heat transport mode across the cell. We identify in the experiments 5 out of the 6 possible modes of heat transport. We highlight the occurrence of brine convection through the mushy ice and of penetrative convection in stably stratified liquid underlying the ice. 
A one-dimensional multi-layer heat flux model built on the known scaling relations of global heat transport in natural convection systems in liquids and porous media is proposed.
It allows, given the measured porosity of the ice, to predict the corresponding ice thickness, in a unified framework.}  
\end{abstract}

\begin{keywords}
\end{keywords}


\section{Introduction}\label{Introduction}
The evolution of sea ice has important impacts on many environmental and geological processes as well as human activities. Examples include ocean circulation~\citep{clark1999northern,straneo2013north,joughin2012ice,hanna2013ice,stevens2020ocean}, global sea-level rise~\citep{wadhams2004ocean}, land-surface albedo~\citep{curry1995sea,perovich2002seasonal,scagliarini2020modelling}, biodiversity~\citep{post2013ecological}, microplastic dispersion and sequestration~\citep{peeken2018arctic,obbard2014global}, and winter navigation at high latitudes and polar areas~\citep{de2018review,ho2010implications}. Generally, sea ice evolution is associated with the complex fluid dynamic processes occurring in the oceans
which involve wide ranges of spatial and temporal scales. Their descriptions through numerical models in the geophysical context make use of so-called, micro-scale physical parametrizations, that have a key role in the reliability and accuracy of the resulting predictions~\citep{rae2015development,rousset2015louvain}.

In this study, we focus on the role of buoyancy-driven natural convection in determining upper surface sea ice growth and the attainment of dynamical equilibrium. We do this by performing experiments at the laboratory scale in a highly-controllable setting, the Rayleigh-B\'enard cell (see below). With this approach
we aim at providing a highly reliable small-scale parameterization of the process that is crucial for geophysical scale computational models in the wide range of applications mentioned above.

The Rayleigh-B\'enard (RB) system, a fluid layer parallelly confined between a warmer horizontal bottom plate and a cooler top plate and insulated on the lateral sides~\citep{ahlers2009heat,lohse2010small,chilla2012new}, is a paradigmatic model system in the physics of non-linear systems and fluid mechanics. Its behaviour has been studied for decades and it has been crucial in developing the understanding of hydrodynamic instabilities, transition to chaos, turbulence and turbulent transport. Its local and global properties across its wide spectrum of regimes are nowadays well-characterized~\citep{ahlers2009heat,chilla2012new}.
More recently the RB system
has been proven to be also convenient for studying the couplings between liquid-solid phase change and fluid flow phenomena~\citep{esfahani2018basal,favier2019rayleigh}. It has a laboratory scale, it is closed and in/out energy fluxes can be measured, hence highly controllable, yet able to reproduce the rich process of water freezing in the presence of unsteady and turbulent flow conditions. Several recent studies have investigated the influence of temperature~\citep{wang2021growth,yang2023jfm}, inclination~\citep{wang2021ice,yang2022abrupt}, rotation~\citep{ravichandran2022combined}, initial conditions, and aspect ratio~\citep{wang2021equilibrium} on freshwater freezing/melting in the RB system. However, the presence of dissolved salt in sea water adds to the complexity of the processes. In particular two factors need to be considered to properly study the evolution of sea ice.

The first is the mushy structure of sea ice~\citep{feltham2006sea}, which allows for the occurrence of interstitial fluid motion, i.e., brine drainage and thermal convection across the sea ice. Various studies have discussed the mushy structure forming in the solidification of aqueous solutions. The focus has mainly been on onset conditions for convective brine drainage~\citep{wettlaufer1997natural}, mushy phase morphology~\citep{peppin2008steady}, ice layer growth rate~\citep{kerr1990solidification}, ice porosity~\citep{wettlaufer1997natural}, flow structure within the mushy phase~\citep{chen1995experimental,worster1997natural}, and salinity in the liquid phase~\citep{wettlaufer1997natural,peppin2008steady}. While these studies together with relevant reviews~\citep{worster1997convection,anderson2020convective,wells2019mushy} provide a detailed, thorough, and multi-perspective understanding of brine drainage and mushy phase structure, the interactions between the mushy structure and a neighbouring vigorous thermal convective flow are much less explored. A few studies discussing the growth of sea ice either focus on establishing governing momentum and energy equations suitable for numerical simulations~\citep{feltham2006sea,covello2016multiphase,wells2019mushy}, or on theoretical models which view ice growth as purely diffusion-controlled~\citep{worster2015sea,notz2005thermodynamic}. 

\textcolor{black}{The second factor is salty water's peculiar equation of state (EOS), in particular its density dependence on salinity and temperature~\citep{gebhart1977new} and its freezing point depression~\citep{bodnar1993revised}. 
While for normal sea water 
with 3.5\% salinity (i.e., the mass concentration of dissolved salt) the density monotonically decreases as the temperature is increased from its freezing point. This is not the case for less salty water (encountered for instance near river estuaries) or fresh water, where a maximal density 
occurs in the interval $[0,4]^{\circ}$C~\citep{pawlowicz2015absolute,lyman1940composition}.} Salty water's peculiar density dependence may lead to different combinations of stable/unstable stratifications of the liquid phase in the RB cell. In a similar yet simpler scenario, the combination of stable stratification and unstably stratified turbulence has been found to have a significant influence on the freezing dynamics and equilibrium ice thickness of fresh water ~\citep{townsend1964natural, wang2021growth}. Other studies point out that the kind of stratification, either stable or unstable, in touch with the melting front has a key role in determining the melting rate of a pure-ice layer in aqueous solutions~\citep{rosevear2021role,mondal2019ablation,sugawara2000effect,yang2023arxiv}. \textcolor{black}{However, these studies focus on stratifications due to the salinity gradient in double-diffusive convection with linearized EOS salt-water mixtures, i.e. neglecting the above-mentioned important nonlinear dependencies of density with respect to temperature and salinity.}

In this work, we investigate how ice growth and fluid dynamics in salty water couple and co-evolve until the equilibrium in the RB natural convection system is reached. We aim to reveal how sea ice growth rate depends on controlled environmental conditions through its interactions with mushy phase convection and with the variety of density stratification patterns and fluid flows occurring in the underlying liquid phase. 

\textcolor{black}{The paper is organized as follows. §\ref{Experiments} describes the experimental setup and the measurement methodology. §\ref{Mushy phase morphology and growth dynamics} describes the experimental results focusing on the growth dynamics and equilibrium morphology of the ice layer. To provide a first-step understanding of the complex process, in §\ref{Global properties of the equilibrium state} we discuss how the equilibrium ice thickness, its porosity and the saturation time depend on temperature and salinity conditions, as well as the correlation among these global properties. §\ref{Multi-layer heat flux model} introduces a one-dimensional multi-layer heat flux model which describes the equilibrium ice thickness's dependence on the system control parameters and identifies 5 out of 6 possible modes of heat transport in the experimental system. Finally, we end with conclusions and outlook in §\ref{Conclusions and outlook}.}
 
\section{Experiments}\label{Experiments}
\begin{figure}
	\centering
	\includegraphics[width = 1\textwidth]{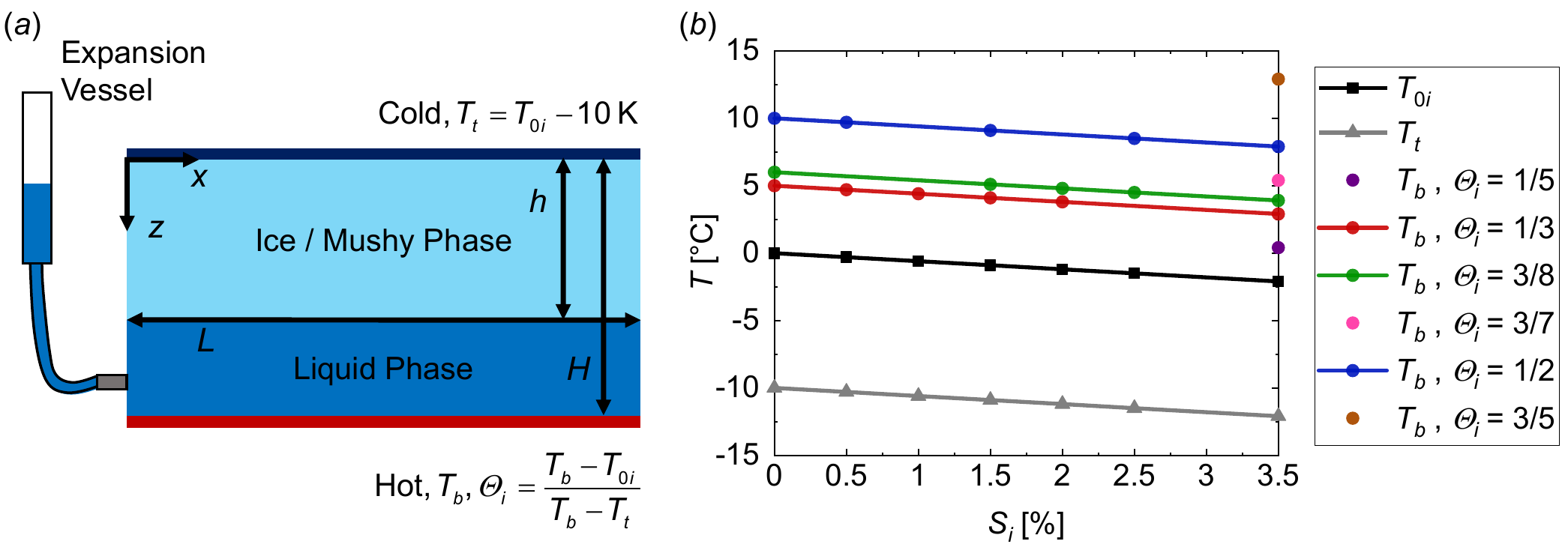}
	\caption{
		\textcolor{black}{Sketch of experimental system and temperature-salinity parameter space.} 
  (a) Cartoon of the experimental setup. The height of the cell is $H=0.12$ m, the aspect ratio is $\Gamma = L/H = 2$, and the width-height ratio is $W/H = 0.5$. An expansion vessel is attached to compensate for the volume change and to monitor the mushy phase growth. The bottom plate temperature is kept at $T_b$ and the top plate temperature $T_t$ is maintained at 10 K below the freezing point $T_{0i}$ at the initial salinity $S_i$. A dimensionless bottom superheat parameter is defined as $\Theta_i=(T_b-T_{0i})/(T_b-T_t)$. \textcolor{black}{(b) Temperature-Salinity parameter space. In this work, we test $\Theta_i =$ 1/5 (purple), 1/3 (red), 3/8 (green), 3/7 (pink), 1/2 (blue), and 3/5 (brown). The initial salinity $S_i$ varies from 0 to 3.5\%. $T_b$ (dots) is shown for each case. $T_{0i}$ (back squares) and $T_t$ (gray triangles) are also plotted as functions of $S_i$.}
	}
\label{Fig1} 
\end{figure}

The experiments are performed in a rectangular cuboid RB cell (height $H=0.12$ m, aspect ratio $\Gamma = L/H = 2$, width-height ratio $W/H = 0.5$, i.e., a so-called quasi-two-dimensional system).  An expansion vessel is attached to compensate for the volume change during the freezing process and so to monitor the mushy phase growth, see figure~\ref{Fig1}(a). The top and bottom plate temperatures are measured with thermistors and controlled with circulating baths. Sidewalls are isolated with foam material to approach ideal adiabatic boundary conditions. 
\textcolor{black}{The cell is initially filled with degassed salty water of initial salinity $S_i$. Here we use sodium chloride (NaCl), the principal solute of sea water~\citep{lyman1940composition}, as the salt.}
$S_i$ is varied from 0 (fresh water) to 3.5\% (average ocean salinity) with the corresponding freezing point $T_{0i}$~\citep{hall1988freezing} varying from 0 to -2.1 $^{\circ}$C. The (hot) bottom plate temperature, $T_b$, which is equal to the initial liquid temperature, ranges from 0.4 to 12.9 $^{\circ}$C, and the (cold) top plate temperature $T_t$ is set 10 K below the freezing temperature, hence $T_t = T_{0i}-10$ K (from -10 to -12.1 $^{\circ}$C depending on $S_i$). A dimensionless bottom super-heat parameter can then be defined as $\Theta_i = (T_b-T_{0i})/(T_b-T_t)$. 
\textcolor{black}{This parameter characterizes the initial ratio of the admissible temperature difference in the liquid phase over the total gap across the cell. It can also slightly evolve during the experiment. This is because the melting temperature depends on the actual salinity, and at the equilibrium state $S_e$ and so $T_{0e}$ might differ from the initial ones (actually $S_e$ can be as high as $7\%$ with $T_{0e}\approx-4.2$ $^{\circ}$C, see figure~\ref{Fig5}b).} Since the total quantity of solute is constant, one expects an enhancement of salinity in the liquid due to salt ejection from the mush (and so an increase of $\Theta$).

 In this work, we first test the states $\Theta_i=$ 1/3 (red), 3/8 (green) and 1/2 (blue) with salinities $S_i \in [0, 3.5]\%$. Cases with $\Theta_i =$ 1/5 (purple), 3/7 (pink), 3/5 (brown) and $S_i = 3.5$\% are also performed to test the results in a wider temperature range. 
 \textcolor{black}{A visualization of the temperature-salinity parameter space explored in the experiments is provided in figure~\ref{Fig1}(b). It reports the temperature values  $T_b$ (dots), $T_{0i}$ (black squares) and $T_t$ (gray triangles) for different $\Theta_i$ and $S_i$.} 
In this experimental configuration, the ice/mushy phase grows from the top plate until the system reaches an equilibrium. We assume that this state is attained when the mean mushy phase thickness $h(t)$ (see figure~\ref{Fig1}a) changes less than 1\% in eight hours, with the maximum experiment duration being three days. The ice or mushy phase thickness and its porosity $\phi(t)$ (i.e. the volume fraction of residual liquid in the ice or mushy phase) can be obtained respectively by identifying the ice-liquid interface in photos of the system, and by using the expansion vessel readings together with the assumption of total mass conservation in the system. 
\textcolor{black}{The evaporation from the expansion vessel is negligible as it only causes a mass loss less than 0.1\% of the total mass of the system.} We refer to Appendix \ref{Data processing} for more details on the measurement protocols.

\section{Mushy phase morphology and growth dynamics}\label{Mushy phase morphology and growth dynamics}
\begin{figure}
	\centering
	\includegraphics[width = 1\textwidth]{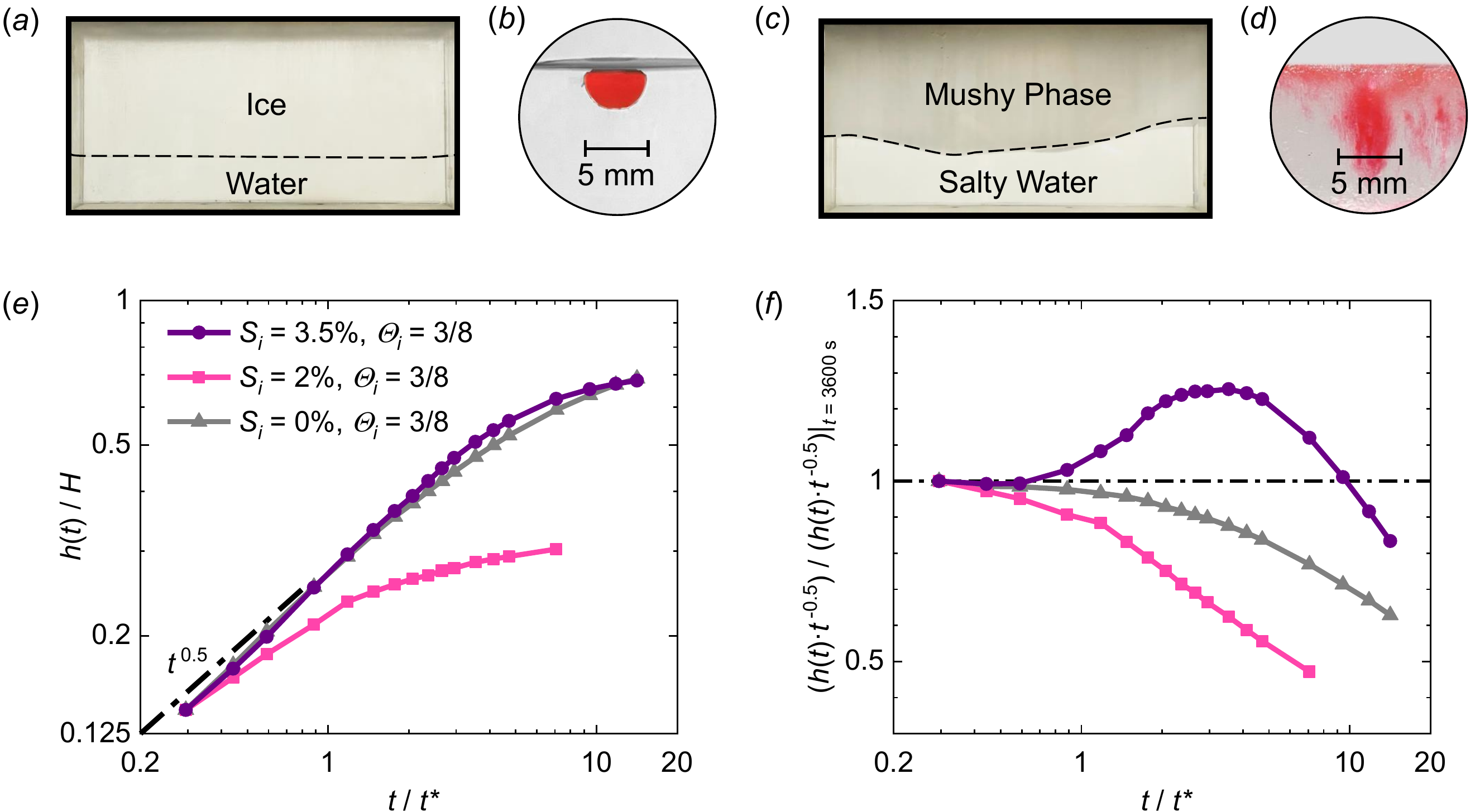}
	\caption{
		Images of the iced layer together with its internal structure, and evolution of the freezing front. (a-d), Photos of ice/mushy phases without (a,b) and with (c,d) porosity. The experimental conditions for (a,b) are $S_i=0\%$ for (c,d) $S_i=3.5\%$, both with $\Theta_i = 3/8$.
		In (b,d), the red dye of the same amount and concentration is deposed into a small hemispherical pit carved at the same location on the upper surface of the corresponding solid matrix in (a,c). The ice in (a,b) is transparent and the dye does not spread. Mushy phase with porosity in (c,d) is relatively opaque and the dye spreads anisotropically, possibly indicating the presence of channels. 
		(e) Dynamics of the freezing front. The mushy phase thicknesses $h(t)$ non-dimensionalized by $H$ are plotted as a function of time non-dimensionalized by the diffusive time scale $t^* =H^2/\kappa_{ice}=1.22\times 10^4$ s (where $\kappa_{ice}$ is ice thermal diffusivity)  for three experimental runs. (f) The same data as that in (e) are compensated by the diffusive behaviour, $h(t)/\sqrt{t}$ (normalized by the first data to compare different cases).  
	}
\label{Fig2} 
\end{figure}

Figures~\ref{Fig2}(a),(c) show photos of the experiment at its equilibrium state in two typical cases ($S_i=0$, 3.5\% and $\Theta_i=3/8$). One can readily appreciate the different thicknesses of the iced domains (thinner in the salty-water case) and the different interface shapes between the two cases. However, their dissimilarity is not only volumetric and morphological but also structural.
To illustrate this we hollowed out a tiny (5 mm) hemispherical cavity on the top horizontal surface of the ice matrix on which we deposed a drop of red dye. 
\textcolor{black}{Figures~\ref{Fig2}(b),(d) show two lateral view photos of the stain after 5 minutes to form its deposition.}
We observe that when $S_i=0$ (panel b) the iced phase is transparent and the dye does not spread into it, indicating the dense structure of the matrix. On the contrary, at  $S_i=3.5\%$ (panel d) the iced phase looks opaque and the dye spreads anisotropically, this highlights a complex porous medium filled with liquid (i.e., a mushy phase structure) with internal vertical channels. 

Noteworthy, the different iced phase structures correspond to different freezing dynamics. This is well illustrated in figures~\ref{Fig2}(e),(f), where the temporal growth of the freezing front, $h(t)$, is shown for three sets of experimental conditions corresponding to pure water, intermediate salinity and seawater salinity, all at $\Theta_i=3/8$. Though all three cases display an initial period of diffusive ice phase growth, i.e., $h \propto \sqrt{t}$, later on the growth rate decreases for the case $S_i=0$ (gray triangles), 2\% (pink squares); while when $S_i=3.5\%$ (purple dots) the instantaneous growth-rate increases and even exceeds diffusive growth rate before reaching the equilibrium. It is remarkable also to note that the equilibrium height is not simply proportional to the salinity.  
The above highlighted temporal evolution of the ice growth in a salty solution depends on the internal structure of the mushy phase that in turn might also change in time (e.g. due to the formation of brine channels~\citep{wettlaufer1997natural,worster1997natural}). A quantitative description of this process is challenging because it requires detailed knowledge of the mushy microstructure. The same is true for any attempt to characterize the morphology of the solid-fluid interface. For this reason, in the following we focus on a simpler question, we try to tackle the globally averaged equilibrium state that is reached asymptotically in time. In other words, we aim at understanding the factors that control the average thickness, denoted $h_e$, of the iced mushy phase.

\begin{figure}
	\centering
	\includegraphics[width = 0.9\textwidth]{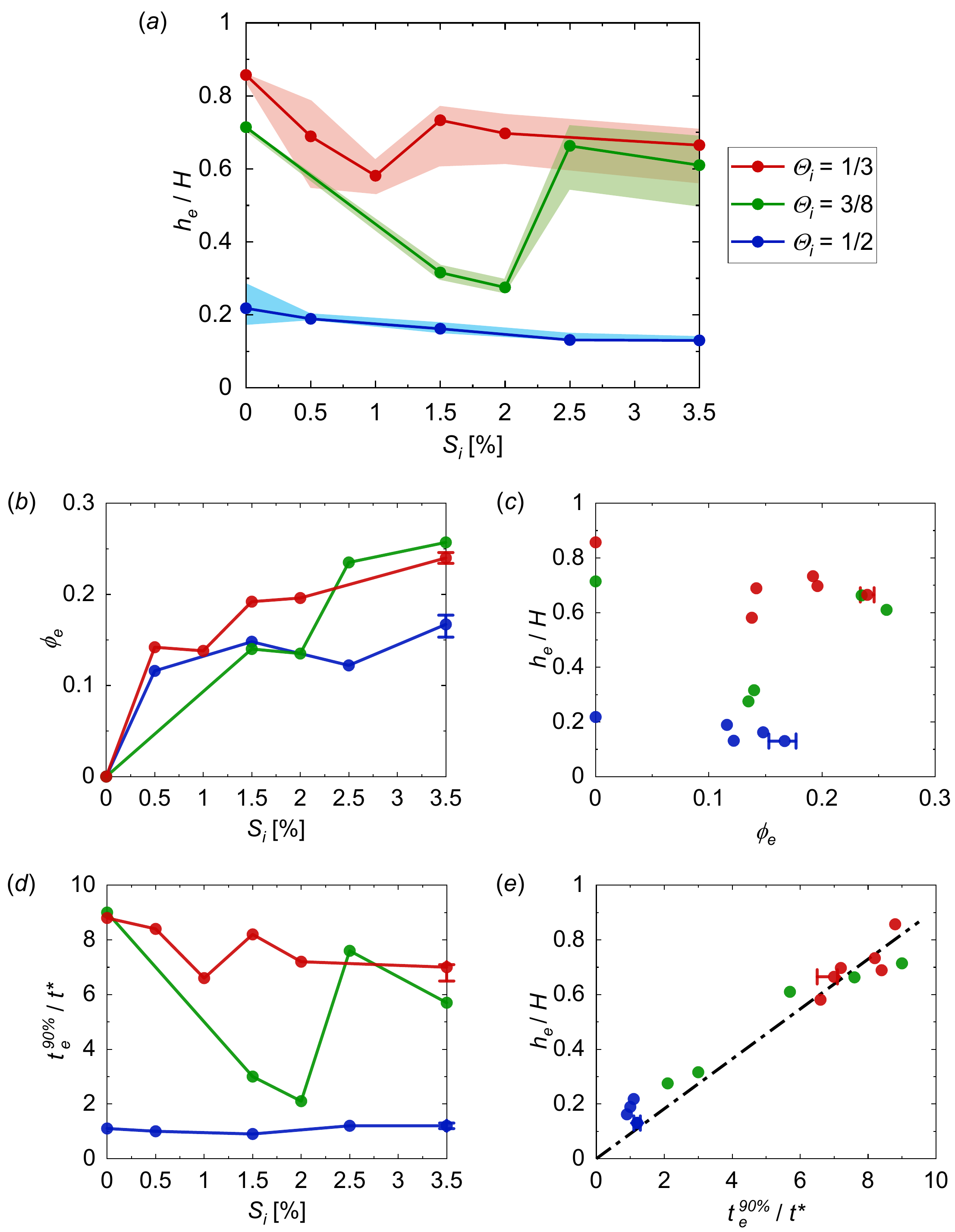}
	\caption{
		Global properties of the equilibrium state from experiments. The initial superheats are respectively $\Theta_i =$ 1/3 (red), 3/8 (green), and 1/2 (blue), and the intial salinity $S_i$ varies from 0 to $3.5 \%$. The error bars are measured by repeated experiments in two cases with $S_i = 3.5\%$ and $\Theta_i = 1/3$, 1/2 respectively. (a) The mean mushy phase thickness $h_e$ (dots) initially decreases with $S_i$ for all three superheats, and it is followed by a sharp increase for the $\Theta_i = 1/3$ and $3/8$ cases. The shaded areas indicate the minimal and maximal spatial variations of mushy phase thicknesses. The thicknesses are non-dimensionalized by $H$. 
  \textcolor{black}{(b) The mean mushy phase porosity $\phi_e$ (squares) exhibits a non-monotonous yet overall increasing trend with $S_i$. (c) $h_e$ does not exhibit a statistically strong correlation with $\phi_e$.}
  (d) The saturation time $t_e^{90\%}$ (triangles) is defined as the time to reach 90\% of the equilibrium mushy phase thickness $h_e$ and non-dimensionalized by the diffusive time scale $t^*=1.22\times 10^4 s$. The variations of $t_{e}^{90\%}$ with $S_i$ are consistent with those of $h_e$. (e) $h_e$ shows a clear positive correlation with $t_{e}^{90\%}$. 
  \textcolor{black}{A linear fitting gives $h_e/H = 0.0912 \cdot t_{e}^{90\%}/t^*$ (dash-dot line), with the correlation coefficient $R^2=0.9846$.}
     }
	\label{Fig3} 
\end{figure}

\section{Global properties of the equilibrium state}\label{Global properties of the equilibrium state}

We display in Figure~\ref{Fig3} two important spatially averaged properties of the mushy phase at the equilibrium (from now on denoted with a subscript $e$): they are the dimensionless thickness $h_e/H$  and the porosity $\phi_e$. Besides this we report the saturation time (or time to equilibrium) denoted as $t_{e}^{90\%}$, here for practical reasons defined as the time it takes to reach 90\% of the equilibrium thickness non-dimensionalized by the diffusive time scale $t^*=H^2/\kappa_{ice}=1.22\times 10^4$ s. 
\textcolor{black}{The dependence of the thickness $h_e$ as a function of the initial salinity $S_i$ is very different for distinct values of the control parameter $\Theta_i$} (see figure~\ref{Fig3}a): $h_e$ exhibits a similar non-monotonic (decrease-increase-decrease) pattern for $\Theta_i = 1/3, 3/8$ and monotonically decreasing for $\Theta_i = 1/2$. \textcolor{black}{These trends are in part reflected on the $\phi_e(S_i)$ dependence, which presents considerable fluctuations for intermediate salinity values although its overall trend is increasing for all $\Theta_i$ cases (see figure~\ref{Fig3}b).} 

\textcolor{black}{The fact that a more porous ice matrix (high $\phi_e$) can release more salt into the liquid and as a consequence lower the final freezing point of the solution, $T_{0e}$, hence reducing the size of $h_e$ might give a rational for the globally observed trends. However, this alone is not enough to explain any of the observed fluctuations or non-monotonous dependencies in $h_e$ and $\phi_e$. Moreover, it is here important to notice that the existence of a correlation between $\phi_e$ and $h_e$ is not evident from the present measurements (see figure~\ref{Fig3}c). 
If, for instance, mass conservation and geometry were simply ruling the relation between the equilibrium mushy layer thickness and the mushy layer porosity, then one would have $h_e(1-\phi_e) = const.$ (where the constant represents the height ice-layer in the case zero salinity and correspondingly no-porosity). This implies an increasing trend of $h_e$ with $\phi_e$, while figure~\ref{Fig3}(c) suggests just the opposite (one can notice that a decreasing $h_e(\phi_e)$ for fixed values of $\Theta_i$). We conclude that the $h_e(\phi_e$) relation is a non-trivial function of the system control parameters $S_i$ and $\Theta_i$.}
 
 \textcolor{black}{Finally, we observe that the trends for $t_{e}^{90\%}$ are instead consistent with those of $h_e$ (figure~\ref{Fig3}d) and a clear linear correlation can be seen among the two quantities (figure~\ref{Fig3}e).}
 This correspondence can provide a way to estimate the time to equilibrium just by focusing on the magnitude of $h_e$. Therefore, in the following we will focus on the modelling of the physical mechanisms responsible for the variations of $h_e$. The question we wish to address is: Can the equilibrium global properties be understood in terms of a model that considers just (i) the role of porosity (ii) the role of thermal stratification, on the steady state of the heat transfer process in the system without taking into account its complex transient dynamics?

\section{Multi-layer heat flux model}\label{Multi-layer heat flux model}

To answer the above question, we need to understand how $h_e$ and $\phi_e$ are related and how the heat transport across the system is influenced by the convection in the mushy phase and by the combination of stable stratification and unstably stratified turbulence. We will show that a one-dimensional (or integral) multi-layer heat flux model that relies on scaling laws known for natural convection in fluid and porous media can explain $h_e$ with good accuracy once $\phi_e$ is provided as input from the experiment. \textcolor{black}{On the other hand we observe that a quantitative description of $\phi_e$ requires a model for the evolution of the microscopic structure of the porous ice matrix. This is a challenging task that currently goes beyond our capabilities and that we leave for future researches. In summary, the multi-layer heat flux model presented in the following allows to explain the complex relation linking $h_e$ to $\phi_e$, when the latter is considered as an independent variable.}

\begin{figure}
	\centering 
	\includegraphics[width = 1\textwidth]{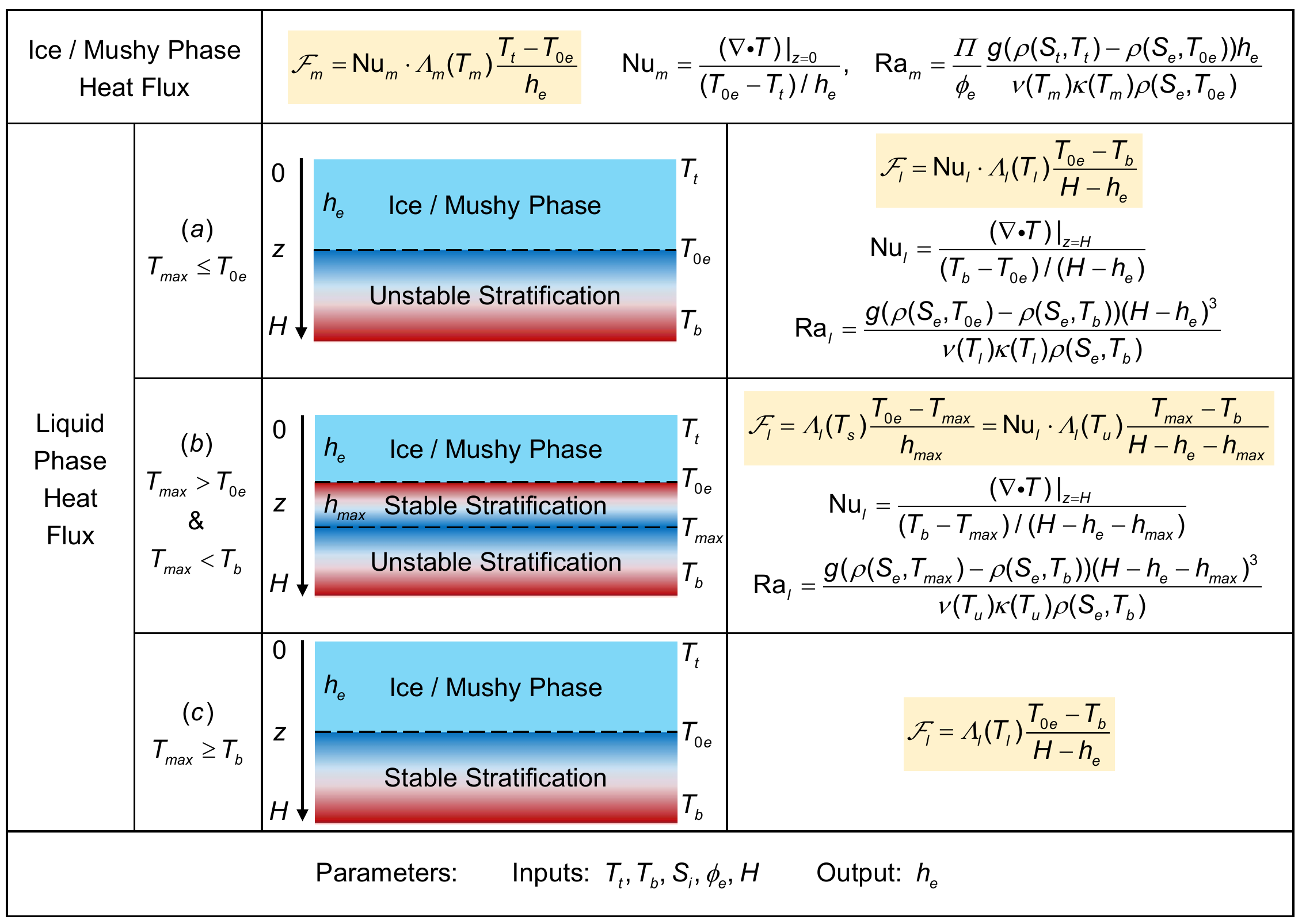}
	\caption{
		\textcolor{black}{Synoptic representation of the averaged multi-layer heat flux model. Row 1 provides the adopted parameterization for the global heat flux expression in the mushy phase, $\mathcal{F}_m$, together with the definitions of Nu$_m$ and Ra$_m$. Row 2-4 report the expression for the global heat flux in the liquid phase, $\mathcal{F}_l$, together with the Nu$_l$ and Ra$_l$ definitions (Column 4). Three different combinations of stratifications may exist, they are: (a) unstable, (b) mixed, (c) unstable. The discriminating criterion for their occurrence is the monotonicity of liquid density within the temperature range $[T_{0e}, T_b]$, i.e., the relative magnitude of $T_{0e}$, $T_b$ and the maximum-density temperature $T_{max}$ (Column 2) which depends on the salinity $S_e$. Column 3 gives the sketches of these combinations, together with the layer thicknesses and the interface temperatures. The color gradient indicates the density gradient in the liquid phase, with red representing lighter liquid and blue representing heavier liquid. The model is supplemented with five input parameters, respectively $T_t$, $T_b$, $S_i$, $\phi_e$ and $H$, and leads to an estimation of $h_e$ (Row 5) through the assumption of a statistically steady heat flux balance among layers.}
	}
	\label{Fig4} 
\end{figure}

We assume that the position of the solid-fluid interface in the system at the equilibrium, $h_e$, is the result of the balance between the mean heat-flux across the mushy-phase, $\mathcal{F}_m$, and the one in the liquid, $\mathcal{F}_l$ (see figure~\ref{Fig4}). Both these quantities require a parameterization, which we detail in the following. For conciseness and better readability, we refer the reader to Table~\ref{tab:table1} in Appendix~\ref{Symbols} for the denomination of the many physical quantities and material properties involved in the model.

Mushy phase: We express the heat flux across the mushy phase in the statistically steady condition in the following form:
\begin{equation} \label{eq:Fm}
	\mathcal{F}_m = \rm Nu \it _m \cdot \rm \Lambda \it_m(T_m)\frac{T_t-T_{0e}}{h_e}.
\end{equation}
Eq. (\ref{eq:Fm}) involves the mushy phase Nusselt number, Nu$_m$, i.e. the ratio between the total and the conductive heat transfer across the mushy phase. Its dependence as a function of the Rayleigh number, Ra$_m$, is known from previous studies on convection in porous media. 
\textcolor{black}{Here we use the parametrization by~\citet{gupta1973bounds}:} 
\begin{equation} \label{eq:Num}
	\rm Nu \it _m=\left\{
	\begin{aligned}
		&1 ,& \rm Ra \it _m < \rm Ra \it _{mcr}, \\
		&1.3338+0.0099\rm Ra \it _m,& \rm Ra \it _m \ge \rm Ra \it _{mcr},
	\end{aligned}
	\right .
\end{equation}
\textcolor{black}{where the mushy-critical Rayleigh number is Ra$_{mcr}=4\pi^2$.}
The Rayleigh number ($Ra_m$) for the flow in a porous medium is defined as:
\begin{equation}\label{eq:Ram}
\rm Ra \it _m=\frac{ \rm \Pi \it}{\phi_e}\frac{g( \rho(S_t, T_t)- \rho(S_e, T_{0e}))h_e}{\nu(T_m)\kappa(T_m) \rho(S_e, T_m)}.
\end{equation}
\textcolor{black}{This definition involves the evaluation of the permeability $\Pi$, a quantity describing the conductivity of a porous medium to fluid flow~\citep{pal2006simple}. The permeability is a function of the structure of the porous ice matrix and is notoriously difficult to be measured, in particular for ice, because it is highly sensitive to the environmental conditions and to the history of the medium. Several previous studies including \citet{kawamura2006measurements}, \citet{petrich2006modelling} and \citet{polashenski2017percolation} revealed that there exists a threshold porosity ($\phi_c$) below which the ice behaves impermeable. Above such threshold 
the permeability is well approximated by a power-law function of the porosity.
We adopt here a similar functional parameterization to relate $\Pi$ to $\phi_e$:}
\begin{equation}\label{eq:permeability}
	\Pi=\left\{
	\begin{aligned}
		&0 ,& \phi_e < \phi_{cr}, \\
		&C \ (\phi_e - \phi_{cr})^\alpha,& \phi_e \ge \phi_{cr},
	\end{aligned}
	\right.
\end{equation}
where the three adjustable parameters are chosen as $C=7 \times 10^{-8}$ m$^2$, $\alpha=3$ (as at the leading order of Carmen-Kozeny relation~\citep{wells2019mushy}) and $\phi_{cr}=0.054$ to best fit the experimentally measured $h_e$. Although these values are ad hoc for our study, the modelled $\Pi$ is of the same order of magnitude as the ones reported in sea-ice laboratory and field researches~\citep{kawamura2006measurements,petrich2006modelling,polashenski2017percolation}.

Liquid phase: Similarly, the mean heat flux in the liquid phase can be expressed in the following approximate form:
\begin{equation}\label{eq:Fl}
\mathcal{F}_l = \left\{
\begin{aligned}
& \rm Nu \it_l \cdot \rm \Lambda \it_l(T_l) \frac{T_{0e}-T_b}{H-h_e},&    T_{max}& \le T_{0e}, \\
&\rm \Lambda \it_l(T_s) \frac{T_{0e}-T_{max}}{h_{max}} =
\rm Nu \it_l \cdot \rm \Lambda \it_l(T_u) \frac{T_{max}-T_b}{H-h_e-h_{max}},& T_{max}&\in (T_{0e},T_b),\\
& \rm \Lambda \it_l(T_l)\frac{T_{0e}-T_b}{H-h_e},& T_{max} &\ge T_b. \\
  \end{aligned}
  \right .
\end{equation}
Note that the above expression is composed of three distinct cases according to the value, $T_{max}$, of the temperature corresponding to the maximal density of the liquid solution at the equilibrium with respect to $[T_{0e}, T_b]$. This reflects in three distinct cases: (a) unstable stratification, (b) a stable layer above an unstable one and (c) a single stable layer, see figure~\ref{Fig4}.
The scale $h_{max}$, appearing in case (b), is the height of the stably stratified layer in the intermediate case. It can be determined by equating the two fluxes across the stable and unstable layers of liquid. The dimensionless heat flux in the liquid is accounted for by the Nusselt number, Nu$_l$, dependence on the Rayleigh number, Ra$_l$, a relation that is relatively well known. 
\textcolor{black}{~\citet{wang2021growth} proposed the parametrization below by fitting the data from their simulations on the freezing of fresh water in an RB convection system, which leads to theoretical predictions agreeing well with their experiment results. The same parametrization is adopted in the present study:}

\begin{equation}\label{eq:Nul}
	\rm Nu \it _l =\left\{
	\begin{aligned}
		&1 , &\rm Ra \it_l &< \rm Ra \it _{lcr}, \\
		&0.12+0.88 \rm Ra \it_l / \rm Ra \it _{lcr}, &\rm Ra \it_l &\in [\rm Ra \it _{lcr}, \rm 1.23\rm Ra \it _{lcr}], \\
		&0.27(\rm Ra \it_l -\rm Ra \it _{lcr} \rm)^{0.27}, &\rm Ra \it_l &> \rm1.23 Ra \it _{lcr},
	\end{aligned}
	\right .
\end{equation}
\textcolor{black}{where the critical Rayleigh number for the onset of convection is Ra$_{lcr}=1708$.}
We note that the definition of the Rayleigh number in the liquid is needed only for the two cases that involve a density unstable stratification (cases (a) and (b) in figure~\ref{Fig4}). We use:
\begin{equation}\label{eq:Ral}
\rm Ra_{l} = \left\{
\begin{aligned}
&\frac{g( \rho(S_e, T_{0e})- \rho(S_e, T_b))( H-h_e )^3 }{\nu(T_l) \kappa(T_l) \rho(S_e, T_l)}, & T_{max}&\le T_{0e}, &\\
&\frac{g( \rho(S_e, T_{max})- \rho(S_e, T_b)( H-h_e-h_{max}) ^3 }{\nu(T_u) \kappa(T_u) \rho(S_e, T_u)},& T_{max}&\in (T_{0e},T_b).&
\end{aligned}
\right .
\end{equation}

Finally, the model tuning requires the numerical values of material properties (see Appendix~\ref{Material properties} for the empirical formulas adopted) which are evaluated at the equilibrium salinity $S_e$. Given the fact that in the present system no salinity source exists, that liquid in the mushy phase only contains a small salt fraction, and that salt is mixed in the liquid either vigorously by convection or slowly by diffusion, we can assume that the $S_e$ is uniform in the liquid layer. 
\textcolor{black}{Hence, $S_e$ can be implicitly related to $h_e$ and the inputs of the model by the mass conservation of the salt:}
\begin{equation}\label{eq:salt}
S_e\rho(S_e, T_{mean})(H-h_e(1-\phi_e))=S_i\rho(S_i, T_{b})H.
\end{equation}
\textcolor{black}{It is worth noticing that Eq. (\ref{eq:salt}) is suitable only for the present system (when applied to other systems, e.g. open systems, a measurement of $S_e$ must be supplied).}
The evaluation of Ra$_m$ requires also the value of  $\rho(S_t, T_t)$. While $T_t$ is known as it is externally imposed, $S_t$ is not. We assume here that the latter takes the value such that $T_t$ is a local freezing point, $T_0(S_t)=T_t$, in other words we assume local thermodynamical equilibrium within the mushy phase~\citep{wells2019mushy}. 
\textcolor{black}{Finally, by simultaneously solving the algebraic equations of heat flux balance $\mathcal{F}_m = \mathcal{F}_l$ and the mass conservation of the salt (\ref{eq:salt}), using the measured value $\phi_e$, the model leads to the equilibrium mushy phase thickness $h_e$. We refer to Appendix \ref{Model operation} for a more detailed explanation of how the model is numerically implemented.}

 \begin{figure}
	\centering
	\includegraphics[width = 1\textwidth]{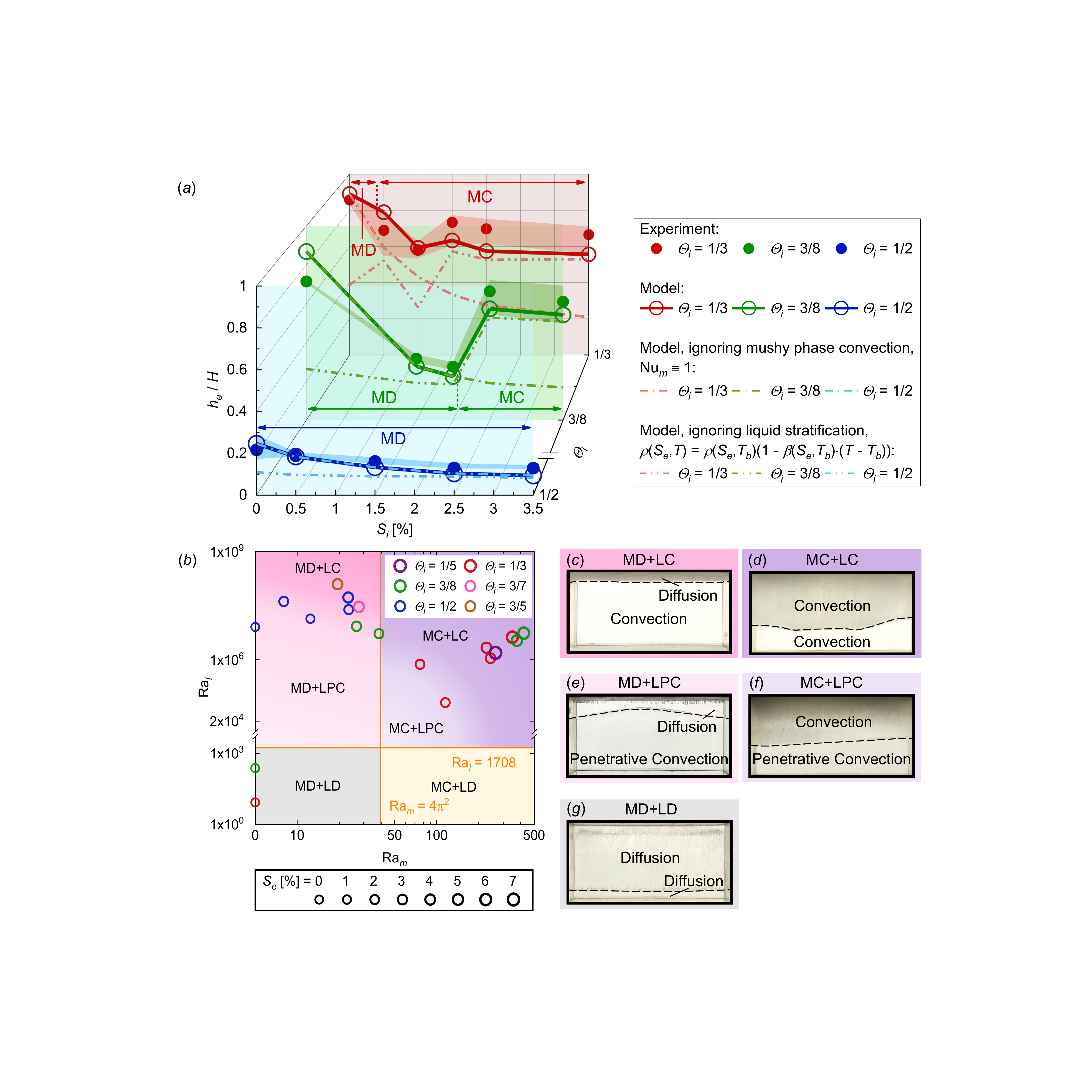}
	\caption{
		Model results for ice/mushy phase thickness compared with the experiments, and identification of heat transport modes. MD/MC stands for diffusive/convective heat transfer in the mushy phase, similarly LD/LC stands for diffusive/convective heat transport in the liquid phase, 
  \textcolor{black}{while LPC stands for diffusive and convective heat transport due to penetrative convection in the liquid phase, i.e., a combination of diffusive and convective heat transport as a stable stratification lies above an unstable one.} (a) The initial superheat values are respectively $\Theta_i = 1/3$ (red), 3/8 (green), and 1/2 (blue), and the initial salinity $S_i$ varies from 0 to $3.5 \%$. The calculated $h_e$ (lines with open circles) is close to the experimental results (dots). The spatial variations of mushy phase thicknesses from the experiments (shaded areas) are also included, together with results of reduced models ignoring mushy phase convection (Nu$_m \equiv 1$, dash-dot lines) and ignoring liquid stratification (linear density dependence on temperature $\rho(S_e, T)=\rho(S_e, T_b)(1-\beta(S_e, T_b)\cdot(T-T_b))$, dash-double-dot lines). (b) Phase-space diagram based on the two effective control parameters, Ra$_m$ and Ra$_l$. 
  \textcolor{black}{The third effective control parameter $S_e$ is not presented as an axis for readability. Instead different circle sizes are used to distinguish the different $S_e$.}
  There are in total 6 heat transport modes of the system. Our experiments fall within 5 cases. Different circle colors are used to distinguish different $\Theta_i$. The highest salinity ($S_e \gtrsim 2.7 \%$) is expected to suppress the LPC regime as it removes the density anomaly of the solution. (c-g) Images from the experiment for typical cases corresponding to the 5 observed heat transport modes.
	}
	\label{Fig5} 
\end{figure}

 \textcolor{black}{The model results are all illustrated in figure~\ref{Fig5}.} Figure~\ref{Fig5}(a) shows the mean mushy phase thickness $h_e/H$ as a function of $S_i$ together with its comparison with the experiment results. At $\Theta_i=1/2$ (blue lines and symbols) where ice thickness decreases for increasing salinity, the agreement with the model is fairly good. The model suggests that in this condition the heat transfer through the mushy phase remains diffusive (mushy diffusion, MD state). Increasing the salinity the freezing point is lowered and this slightly reduces the frozen layer thickness. The role of porosity is marginal as here it stays approximately constant with $S_i$ (see figure~\ref{Fig3}b). For $\Theta_i=1/3$ and $3/8$ (red and green lines and symbols), the non-monotonous behaviour is also captured by the model. 
 
 It is relevant here to also consider two reduced versions of the model, one ignoring the occurrence of convection in the mushy phase (by assuming Nu$_m \equiv 1$, dash-dot lines in figure~\ref{Fig5}a), the other neglecting the non-monotonous density stratification in the liquid phase (by assuming linear density dependence on temperature, i.e., $\rho(S_e, T)=\rho(S_e, T_b)(1-\beta(S_e, T_b)\cdot(T-T_b))$ where $\beta(S_e, T_b)$ is the liquid thermal expansion coefficient at the bottom plate, dash-double-dot lines). The comparison of the calculated $h_e$ stemming from the two reduced models allows to interpret the ``jump'' or sudden increase in the ice thickness as being associated with the occurrence of convection in the mushy phase (mushy convection, MC state) with increased $\phi_e$ (see figure~\ref{Fig3}b). Convection strengthens the heat transport, promoting a larger mushy phase growth rate, because $\textrm{d}h/\textrm{d}t \propto \mathcal{F}_m - \mathcal{F}_l$. 
 \textcolor{black}{The equilibrium mushy layer thickness, as a result of the competitive balance between $\mathcal{F}_m$ and $\mathcal{F}_l$, also increases because of the enhanced heat transfer efficiency in the mushy phase.}
 
   However, the mushy phase convention alone is not sufficient to explain the tendency of $h_e$ (see the dash-double-dot lines in figure~\ref{Fig5}a). Especially large derivations from experiment results can be seen when $S_i$ is low as they fall into LD (liquid diffusion) and LPC (liquid penetrative convection) regimes (see figure~\ref{Fig5}b). In the LD state, only diffusive heat transport occurs in the liquid phase, either because the whole liquid phase is stably stratified or because the liquid phase is too thin to support convection. In the LPC state, there is always a stable stratification in contact with the mushy phase above the unstably stratified turbulence, also known as penetrative convection. The presence of this stable layer limits the below convective motion and prevents the mushy phase from a direct penetration of the convection. As an effect the heat transport across the liquid phase is relatively low, enabling the mushy phase to grow thick. When $S_i$ is increased, the temperature differences decrease in the stably stratified region (from $T_{0e}$ to $T_{max}$) and increase in the unstably stratified turbulence (from $T_{max}$ to $T_b$). As a consequence the convective motion intensifies becoming even more turbulent, leading to strong heat transport across the liquid phase and so decreasing $h_e$. 
   \textcolor{black}{Similarly, the temperature difference in the unstable stratification is smaller when $\Theta_i$ is lower. The unstably stratified turbulence is weaker and the thermal resistance in the stable stratification above plays a more important role in determining the heat transport of the whole liquid phase. This account for the relatively larger errors between the dash-double-dot lines (assuming the entire liquid phase to be unstably stratified) and the experiment results in figure~\ref{Fig5}(a) when $\Theta_i=$ 1/3 and 3/8, compared to cases with $\Theta_i=$ 1/2 and the same $S_i$.}
   By further increasing $S_i$, the stable stratification layer disappears and the heat transport in the liquid phase is purely convective (liquid convection, LC state, see figure~\ref{Fig5}b). 
   
  The remaining discrepancies between the model and the experiments are likely to be related to the modelization of the mushy phase, in particular from neglecting its non-homogeneous internal structure and the uncertainties in the parameterization of its permeability. 
  \textcolor{black}{It is also worth mentioning that the heat transport in the LPC regime can be also clearly described by means of two alternative parameters introduced in~\citet{wang2021universal}. They are the so-called density inversion parameter (which here reads $(T_{max}-T_{0e})/(T_b-T_{0e})$) and the Rayleigh number relative to the whole fluid layer. It is an alternative but equivalent approach with respect to the one adopted here.}
   
  In summary, the possible heat transport modes in the system are 6, depending on whether the heat transport is purely diffusive (MD) or convective (MC) in the mushy phase and whether the heat transport in the liquid phase is purely diffusive (liquid diffusion LD), diffusive and convective in distinct superposed layers (LPC), or convective (LC).
  The occurrence of these states is determined by the magnitude of the two control parameters Ra$_m$ and Ra$_l$. The boundaries between purely diffusive (MD and LD) and convective (MC, LPC and LC) heat transport are the critical Rayleigh numbers, respectively Ra$_{mcr}=4\pi^2$ in the mushy phase and Ra$_{lcr}=1708$ in the liquid.
  On the other hand the transition from LPC to LC is due to the disappearance of temperature anomaly in the equation of state of the salty water solution. This occurs when $T_{max}<T_{0e}$ or equivalently roughly when $S_e \gtrsim 2.7$\%. The experimental measurements can be mapped into this posterior three-dimensional phase space, identified by the triplet (Ra$_m$, Ra$_l$, $S_e$), see figure~\ref{Fig5}(b). 
  \textcolor{black}{It is worth noticing that while $S_i$ varies in $[0,3.5\%]$, $S_e$ can vary in a range from 0\% to about 7\%. We observe that 5 out of 6 possible equilibrium state freezing modes (i.e. the 2 $\times$ 3 combinations of MD,MC states with LD,LPC,LC ones) can be realized in our experiments and identified by our model.}
  Figures~\ref{Fig5}(c-g) show photos of the interface profile in these typical cases. We may observe that non-flat interfaces are always associated with convective motions either in the liquid or the mushy phase or both.

  \section{Conclusions and outlook}  \label{Conclusions and outlook}

We experimentally investigated the complex physical couplings between the freezing of salty water (ice inception, growth and equilibrium) and convective fluid dynamics transport processes. We revealed that the brine convection in the mushy phase and the combination of stable/unstable stratifications and so conductive/convective heat transfer are crucial in the determination of an equilibrium state. The mean mush thickness at the equilibrium can be accurately explained by properly accounting for these two mechanisms via a one-dimensional multi-layer heat flux model which builds on the known global heat transport scaling-law properties in Rayleigh-B\'enard system in liquid and porous media.

The occurrence of convection inside the mushy matrix enhances the heat transport and increases the advancement of the freezing front. On the other hand, the stable stratification weakens convection and heat transport in the liquid phase and sustains a longer diffusive growth of the mushy phase, both leading to a higher mushy thickness. Finally, unstably stratified turbulence promotes convection and heat transfer in the liquid phase, leading to a thinner mushy layer. The resulting heat transport mode at the equilibrium state can be categorized into 6 regimes, depending on whether convective heat transport exists in the mushy region and whether purely diffusive/convective or both regimes of heat transport exist in the liquid phase.

Many open questions remain to be explored. The multi-layer model needs to be further validated and refined in what concerns the modelization of the permeability and the evolution of porosity (which is here taken just as an input from experimental measurements). Furthermore, it would be worthwhile to extend this 1D model to include its temporal behaviour, this seems doable given the fact that its growth rate occurs on time scales that are much longer as compared to the ones of fluid convection; in this sense a quasi-static approximation, as done for pure ice in~\citet{wang2021ice}, might work. However, the comprehension of the morphological mushy-phase interface, the role of brine channels and related desalination calls for more sophisticated spatial models.

The current study provides a first step in the understanding of this complex system. 
\textcolor{black}{It also lays a foundation for further studies investigating the interactions between the detailed mushy phase structure, interface morphology, flow dynamics and heat transport performance of the system, as well as exploring the influences of other factors}
(e.g. the inclination of the system, different geometrical aspect ratio or shapes of the container or the presence of forced next to natural convection). We expect that our findings can lead to a better understanding and global modelling of industrial, geophysical and climatological processes involving flows in liquid and porous media and phase-change in liquid solutions.

\appendix

\section{Data processing}\label{Data processing}

 As the flow is quasi-two-dimensional, the mean mushy phase thickness at the equilibrium $h_e$, as well as the spatial variation of mushy phase thickness, can be estimated with the vertical position of the mushy-liquid interface on the front surface of the experimental cell, which can be obtained by processing the photos of the equilibrium state. The time series  $h(t)$ is obtained with the time series of water level elevation in the expansion vessel $\Delta v(t)$ (non-dimensionalized by $H$):
\begin{equation}
 h(t)=\frac{h_e}{\Delta v_e} \cdot \Delta v(t).
 \end{equation}

The mean mushy phase porosity at the equilibrium $\phi_e$ is obtained by solving mass conservation equations of the system and salt:
\begin{equation}\begin{aligned}
		\rho(S_i, T_b)H&=\rho(S_e, T_{mean})(C_1H-C_1(1-\phi_e)h_e+\Delta v_e)+\rho_{ice}(C_1(1-\phi_e)-C_2)h_e.
\end{aligned}\end{equation}
\begin{equation}\begin{aligned}
	S_i\rho(S_i, T_b)H&=S_e\rho(S_e, T_{mean})(C_1H-C_1(1-\phi_e)h_e+\Delta v_e).
\end{aligned}\end{equation}
where fitting parameters $C_1$ and $C_2$ are adopted to compensate for the slight volume change of the experimental cell and the residual dissolved air in the system. The values are chosen as $C_1=1.0033$ and $C_2=0.0105$ to set $\phi_e=0$ for $S_i=0$ and $\Theta_i=1/3,$ 1. 
\textcolor{black}{The calculated $\phi_e\approx0$ for $S_i=0$ and $\Theta_i=3/8$ indicates the soundness of this treatment.}

\begin{figure}
	\centering
	\includegraphics[width = 1\textwidth]{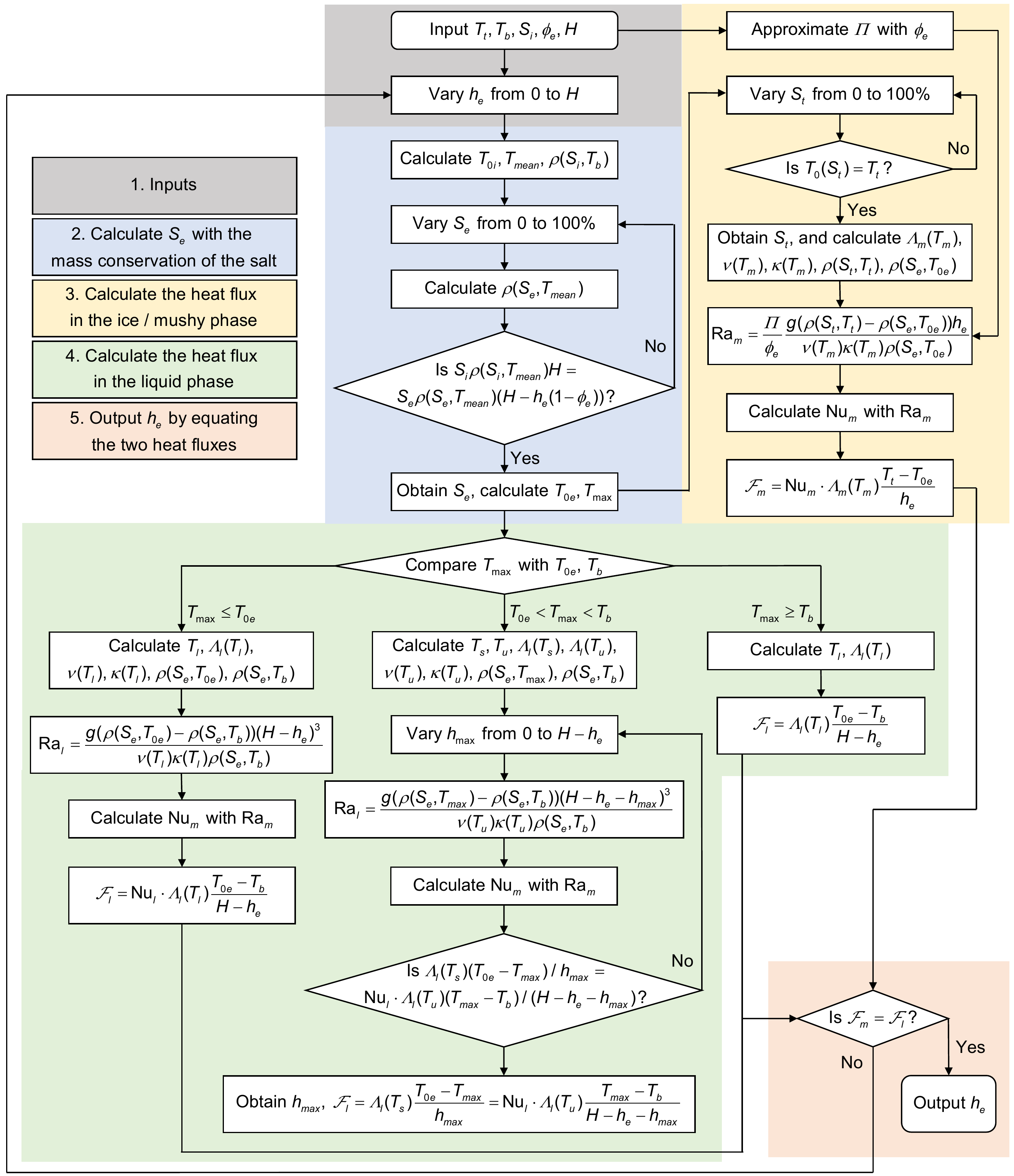}
	\caption{
		Flow chart of the numerical implementation of the model. To obtain the equilibrium mushy phase thickness $h_e$ in the current system, 5 inputs ($T_t$, $T_b$, $S_i$, $\phi_e$ and $H$) are needed, and a hypothetical $h_e$ is assumed (gray blocks). Then the $S_e$ corresponding to the hypothetical $h_e$ is implicitly calculated with the mass conservation of the salt (blue blocks). Next, the permeability $\Pi$ is approximated, the material properties in the ice/mushy phase are evaluated and the heat flux across the ice/mushy phase, $\mathcal{F}_m$, is calculated (yellow blocks). Meanwhile, based on the different combinations of stratifications, the material properties in the liquid phase are evaluated and the heat flux across the liquid phase, $\mathcal{F}_l$, is calculated (green blocks). Finally, the hypothetical $h_e$ is outputted as the real $h_e$ of the system if $\mathcal{F}_m=\mathcal{F}_l$; otherwise the aforementioned process is repeated with a new hypothetical $h_e$ until the optimal $h_e$ is found.
  }
	\label{Fig6} 
\end{figure}

\section{\textcolor{black}{Numerical implementation of the heat-flux model}}\label{Model operation}

\textcolor{black}{The multi-layer heat flux model introduced in §\ref{Multi-layer heat flux model} is implemented via an algorithm summarized in the flow chart of  figure~\ref{Fig6}. It consists of five main steps. First, after inputting $T_t$, $T_b$, $S_i$, $\phi_e$ and $H$, it initializes the ice height to $h_e=0$  (gray blocks). Second, the program calculates the equilibrium salinity $S_e$ corresponding to $h_e$ via the mass conservation of salt relation (\ref{eq:salt}) (blue blocks). Third, the program approximates the permeability $\Pi$ with $\phi_e$ (\ref{eq:permeability}) and evaluates the material properties in the mushy layer, leading to the heat flux across the ice or mushy phase, $\mathcal{F}_m$ (\ref{eq:Fm}) (yellow blocks). Fourth, based on the different stratification patterns (see figure~\ref{Fig4} in §\ref{Multi-layer heat flux model}), the program evaluates the material properties in the liquid phase and calculates the heat flux across the liquid phase (\ref{eq:Fl}) (green blocks). Finally (fifth step), the code examines whether $\mathcal{F}_m=\mathcal{F}_l$. If this is the case, the guessed $h_e$ is outputted, otherwise the program increases $h_e$ by a small step ($10^{-4} h_e/H$) and repeats the aforementioned steps until the convergence of the relation $\mathcal{F}_m=\mathcal{F}_l$.}

\textcolor{black}{As stated in §\ref{Multi-layer heat flux model}, the mass conservation of the salt is an ad-hoc assumption for the present system, which is not applicable in the study of an open system such as the field study of sea ice. In that case, the input $S_i$ and the second step, i.e., the evaluation of $S_e$ can be replaced by the independent measurements of $S_e$ without affecting the model which is established based on thermodynamical balance. We refer to Appendix~\ref{Material properties} for the empirical formulas adopted in our study to evaluate the material properties and table \ref{tab:table1} in Appendix~\ref{Symbols} for all the definitions of the symbols used in this paper.}

\section{Material properties}\label{Material properties}

The multi-layer heat flux model requires the use of material properties, which in general are dependent on temperature $T$ ($^\circ$C) and salinity $S$ (\%). Their parameterizations are described in the following. All properties are in SI units.

1) Freezing point $T_0(S)$~\citep{hall1988freezing}:
\begin{equation}
    T_0(S)=-0.6037S-5.8123\times10^{-4}S^3.
\end{equation}

2) Liquid thermal diffusivity $\kappa(S,T)$:
\begin{equation}
    \kappa(S,T)=\frac{\Lambda_l(S,T)}{\rho(S,T)c_{pl}(S,T)}.
\end{equation}
where $\Lambda_l(S,T)$ is the liquid thermal conductivity (see below), $\rho(S,T)$ is the liquid density (see below), $c_{pl}(S,T)$ is the liquid specific heat capacity. $c_{pl}(S,T)$ can be determined with~\citep{driesner2007system}:

\begin{equation}\begin{aligned}
   c_{pl}(S,T)=a_1c_{pw}(T_{ref}).
\end{aligned}\end{equation}
where $a_1=3.3619-1.6956(x+1.9404)^{0.5}-0.2133x$, $x=\frac{S}{58.443}/(\frac{S}{58.443}+\frac{100-S}{18.015})$ is the mole fraction of salt in the liquid, and $c_{pw}(T_{ref})$ is the heat capacity of pure water at $T_{ref}$. $T_{ref}=a_1T+a_2$, where $a_2=47.8954-32.1103(1-x)-15.7851(1-x)^2$. $c_{pw}(T)$ can be determined with~\citep{chase1998data}: 
\begin{equation}\begin{aligned}
   c_{pw}(T)=&-11302+84.5568T_{abs}-0.1774T_{abs}^2+1.3736\times10^{-4}T_{abs}^3+2.1401\times10^{8}/T_{abs}^2.
\end{aligned}\end{equation}
where $T_{abs}=T+273.15$ is the absolute temperature.

3) Ice thermal diffusivity $\kappa_{ice}(T)$:
\begin{equation}
    \kappa_{ice}(T)=\frac{\Lambda_{ice}(T)}{\rho_{ice}(T)c_{pice}(T)}.
\end{equation}
where $\Lambda_{ice}(T)$ is the ice thermal conductivity (see below), $\rho_{ice}(T)$ is the ice density (see below), $c_{pice}(T)$ is the ice heat capacity. $c_{pice}(T)$ can be determined with~\citep{fukusako1990thermophysical}:
\begin{equation}
     c_{pice}(T)=185+6.89T.
\end{equation}

3) Mushy phase thermal conductivity $\Lambda_m(S,T)$:
\begin{equation}
    \Lambda_m(S,T)=\phi\Lambda_l(S,T)+(1-\phi)\Lambda_{ice}(T)
\end{equation}
where $\Lambda_l(S,T)$ is the ice thermal conductivity (see below), $\Lambda_{ice}(T)$ is the ice thermal conductivity (see below), and $\phi$ is the mushy phase porosity.

4) Ice thermal conductivity $\Lambda_{ice}(T)$~\citep{fukusako1990thermophysical}:
\begin{equation}
    \Lambda_{ice}(T)=2.2156-1.0046\times10^{-2}T+3.4452\times10^{-5}T^2.
\end{equation}

\begin{table}

 \resizebox{1\textwidth}{!}{
\begin{tabular}{ccccc}
\textrm{Symbols}&
\textrm{Physical meanings}&
\tabincell{c}{\\ \\} &
\textrm{Symbols}&
\textrm{Physical meanings}\\

$g$ & gravitational acceleration & \tabincell{c}{\\ \\}
& MD & \tabincell{c}{Only diffusive heat transport exists \\ in the mushy phase} \\

$H$ & Height of the cell &\tabincell{c}{\\ \\}
&Nu$_l$ & Liquid phase Nusselt number\\

$h(t)$ & Mushy phase thickness at time $t$ &\tabincell{c}{\\ \\} 
& Nu$_m$ & Mushy phase Nusselt number\\ 

$h_e$ & \tabincell{c}{Mushy phase thickness at the equilibrium} &\tabincell{c}{\\ \\}  
&Ra$_l$ & Liquid phase Rayleigh number\\ 

$h_{max}$ & \tabincell{c}{Stable stratification thickness at the equilibrium} & \tabincell{c}{\\ \\} 
&Ra$_{lcr}$ & Critical liquid phase Rayleigh number \\ 

$L$ & Length of the cell & \tabincell{c}{\\ \\} 
&Ra$_m$ & Mushy phase Rayleigh number \\

$S_e$ & Equilibrium salinity in the liquid & \tabincell{c}{\\ \\} 
&Ra$_{mcr}$ & Critical mushy phase Rayleigh number \\

$S_i$ & Initial salinity & \tabincell{c}{\\ \\} 
&$\Delta v(t)$ & \tabincell{c}{Water elevation in the expansion vessel \\ at time $t$}\\

$S_t$ & \tabincell{c}{Equilibrium salinity at the top plate,\\ corresponding to freezing point $T_0(S_t)=T_t$} & \tabincell{c}{\\ \\} 
&$\Delta v_e$ & \tabincell{c}{Water elevation in the expansion vessel \\ at the equilibrium}\\

$T_{0e}$ & Equilibrium freezing point at salinity $S_e$ & \tabincell{c}{\\ \\} 
&$\beta(S, T)$ & \tabincell{c}{Liquid thermal expansion coefficient \\ at salinity $S$ and temperature $T$} \\

$T_{0i}$ & Initial freezing point at salinity $S_i$ & \tabincell{c}{\\ \\}
&$\Gamma$ & Aspect ratio of the cell, $L/H$ \\

$T_b$ & Bottom plate temperature & \tabincell{c}{\\ \\}
&$\kappa(T)$ & \tabincell{c}{Liquid thermal diffusivity\\ at salinity $S_e$ and temperature $T$} \\

$T_l$ & \tabincell{c}{Mean temperature of the liquid phase, \\$0.5(T_{0e}+T_b)$}&\tabincell{c}{\\ \\}
&$\kappa_{ice}$ & \tabincell{c}{Ice thermal diffusivity \\ at temperature $T_m$}\\

$T_m$ & \tabincell{c}{Mean temperature of the mushy phase, \\$0.5(T_t+T_{0e})$}& \tabincell{c}{\\ \\}
&$\Lambda_{ice}(T)$ & \tabincell{c}{Ice thermal conductivity at temperature $T$}\\

$T_{max}$ & \tabincell{c}{Temperature where liquid density \\ reaches the maximum at salinity $S_e$} &\tabincell{c}{\\ \\}
&$\Lambda_{l}(T)$ & \tabincell{c}{Liquid thermal conductivity \\ at salinity $S_e$ and temperature $T$}\\

$T_{mean}$ & \tabincell{c}{Mean temperature of the whole system,\\ $0.5(T_t+T_b)$} & \tabincell{c}{\\ \\}
&$\Lambda_m(T)$ & \tabincell{c}{Mushy phase thermal conductivity at salinity $S_e$ \\ and temperature $T$, $\phi_e \Lambda_{l}(T)+(1-\phi_e) \Lambda_{ice}(T)$} \\

$T_s$ & \tabincell{c}{Mean temperature of the stable stratification, \\ $0.5(T_{0e}+T_{max})$} & \tabincell{c}{\\ \\} 
&$\nu(T)$ & \tabincell{c}{Liquid kinetic viscosity\\ at salinity $S_e$ and temperature $T$} \\

$T_t$ & Top plate temperature & \tabincell{c}{\\ \\} 
&$\phi_e$ & Porosity at the equilibrium \\

$T_u$ & \tabincell{c}{Mean temperature of the unstable stratification, \\$0.5(T_{0e}+T_{max})$}&\tabincell{c}{\\ \\} 
& $\phi_{cr}$ & Critical porosity \\

$t$ & Time & \tabincell{c}{\\ \\} 
&$\Pi$ & Permeability \\

$t^*$ & Diffusive time scale, $H^2/\kappa_{ice}$ & \tabincell{c}{\\ \\} 
&$\rho(S, T)$ & \tabincell{c}{liquid density \\at salinity $S$ and temperature $T$}\\

$t_{e}^{90 \%}$ & \tabincell{c}{Saturation time, \\ the time it takes to reach 90\% of $h_e$} &\tabincell{c}{\\ \\} 
& $\rho_{ice}$ & Ice density at temperature $T_m$\\

$W$ & Width of the cell & \tabincell{c}{\\ \\} 
&$\rho_t$ & Equilibrium liquid density at the top plate\\

LC & \tabincell{c}{Only convective heat transport exists \\ in the liquid phase} &\tabincell{c}{\\ \\} 
& $\Theta_i$ & \tabincell{c}{Dimensionless bottom super-heat,\\ $(T_b-T_{0i})/(T_b-T_t)$}\\

LD & \tabincell{c}{Purely diffusive heat transport exists \\ in the liquid phase} &\tabincell{c}{\\ \\} 
&$\mathcal{F}_l$ & Mean heat flux across the liquid phase\\

MC & \tabincell{c}{Convective heat transport exists \\ in the mushy phase} &\tabincell{c}{\\ \\} 
&$\mathcal{F}_m$ & Mean heat flux across the mushy phase\\

\end{tabular}}
\caption{\label{tab:table1}
Symbols used in this paper.
}
\end{table}

5) Liquid thermal conductivity $\Lambda_l(S,T)$~\citep{yusufova1975thermal}:
\begin{equation}\begin{aligned}
    \Lambda_l(S,T)=\Lambda_w(T)(1&-(2.3434\times10^{-3}-7.924\times10^{-6}T_{abs}+3.924\times10^{-8}T_{abs}^2)S\\&+(1.05\times10^{-5}-2\times10^{-8}T_{abs}+1.2\times10^{-10}T_{abs})S^2).
\end{aligned}\end{equation}
where $\Lambda_w(T)$ is the thermal conductivity of pure water at $T$, $T_{abs}$ is the absolute temperature. $\Lambda_w(T)$ can be determined with~\citep{ramires1995standard}:
\begin{equation}\begin{aligned}
   \Lambda_w(T)=-0.9003+2.5006\frac{T_{abs}}{298.15}-0.9938(\frac{T_{abs}}{298.15})^2.
\end{aligned}\end{equation}

6) Liquid kinetic viscosity $\nu(S,T)$:
\begin{equation}
    \nu(S,T)=\frac{\mu(S,T)}{\rho(S,T)}.
\end{equation}
where $\mu(S,T)$ is the dynamic viscosity and $\rho(S,T)$ is the liquid density (see below). $\mu(S,T)$ can be determined with~\citep{simion2015mathematical}:
\begin{equation}\begin{aligned}
   \mu(S,T) &=1.257\times10^{-4}+1.265\times10^{-3}e^{-0.04297T}-1.105\times10^{-3}e^{0.3710x}\\&+2.045\times10^{-4}e^{-0.4231(0.01T+x)}+1.309\times10^{-3}e^{-0.3260(0.01T-x)}.
\end{aligned}\end{equation}
where $x$ is the mole fraction of salt in the liquid.

7) Liquid density $\rho(S,T)$~\citep{gebhart1977new}:
\begin{equation}
    \rho(S,T)=b_1(1-b_2|T-b_3|^{1.895}).
\end{equation}
where $b_1=999.972(1+8.046\times10^{-3}S)$ is the maximum density, $b_2=9.297\times10^{-6}(1-0.02839S)$, $b_3=3.98(1-0.5266S)$ is the temperature where liquid density reaches the maximum.

8) Ice density $\rho_{ice}(T)$~\citep{fukusako1990thermophysical}:
\begin{equation}
    \rho_{ice}(T)=917(1-1.17\times10^{-4}T).
\end{equation}

\section{Symbols}\label{Symbols}

The symbols used in this paper are summarized in TABLE~\ref{tab:table1}.

\section*{Acknowledgements}

We thank Yantao Yang for insightful discussions. This work was supported by Natural Science Foundation of China under grant no. 11988102 and 91852202, and the Tencent Foundation through the XPLORER PRIZE.

\section*{Declaration of Interests}

The authors report no conflict of interest.


\end{document}